\RequirePackage{amsmath}
\documentclass[runningheads,a4paper]{llncs}

\usepackage[utf8]{inputenc} 
\usepackage{xcolor}
\usepackage[color=blue!20, linecolor=blue!80!black]{todonotes}
\usepackage[english]{babel}
\usepackage[utf8]{inputenc}
\DeclareUnicodeCharacter{00A0}{ }
\usepackage{hyperref}
\usepackage{xspace}
\usepackage{amsmath} 
\usepackage[ruled,vlined]{algorithm2e}
\usepackage[capitalise,nameinlink]{cleveref}
\usepackage[binary,squaren]{SIunits} 
\usepackage{amssymb}
\usepackage{listings}
\usepackage{dsfont}
\usepackage{multicol}
\usepackage{cite}
\usepackage{float}
\usepackage{booktabs}
\usepackage{cleveref}
\usepackage{pgfplots}
\usepackage{caption}
\usepackage{subcaption}
\usepackage{soul}
\usepackage{mathtools}
\usepackage{xparse}
\usepackage[inline]{enumitem}
\usepackage{graphicx}
\usepackage{paralist}

\usepackage{booktabs} 
\usepackage{xspace} 
\usepackage{lstautogobble}  

\usepackage{underscore}
\usepackage{enumitem}
\usepackage{pgfplots}
\usepgfplotslibrary{colorbrewer}
\pgfplotsset{compat = 1.15, cycle list/Set1-8} 
\usetikzlibrary{pgfplots.statistics, pgfplots.colorbrewer} 
\usepackage{pgfplotstable}
\usepackage{filecontents}
\usepackage[belowskip=-15pt,aboveskip=0pt]{caption}

\usetikzlibrary{shapes, arrows, positioning, calc}
\makeatletter
\newcommand*{\etal}{%
  \@ifnextchar{.}%
  {et~al}%
  {et~al.\@\xspace}%
}

\makeatother

\newcommand\donotshow[1]{}

\NewDocumentCommand{\twopartdef}{ m m m o}{
  \left\{
  \begin{array}{ll}
    #1 & \mbox{if } #2                                     \\
    #3 & \IfNoValueTF{#4}{\text{otherwise}}{\mbox{if } #4}
  \end{array}
  \right.
}
\NewDocumentCommand{\threepartdef}{m m m m m o}{
  \left\{
  \begin{array}{lll}
    #1 & \mbox{if } #2                                     \\
    #3 & \mbox{if } #4                                     \\
    #5 & \IfNoValueTF{#6}{\text{otherwise}}{\mbox{if } #6}
  \end{array}
  \right.
}
\NewDocumentCommand{\longthreepartdef}{m m m m m m m o}{
  \left\{
  \begin{array}{lll}
    #1 & \mbox{if } #2                                     \\
       & #3                                                \\
    #4 & \mbox{if } #5                                     \\
       & #6                                                \\
    #7 & \IfNoValueTF{#8}{\text{otherwise}}{\mbox{if } #8}
  \end{array}
  \right.
}
\NewDocumentCommand{\fourpartdef}{m m m m m m m o}{
  \left\{
  \begin{array}{llll}
    #1 & \mbox{if } #2                                     \\
    #3 & \mbox{if } #4                                     \\
    #5 & \mbox{if } #6                                     \\
    #7 & \IfNoValueTF{#8}{\text{otherwise}}{\mbox{if } #8}
  \end{array}
  \right.
}

\newcommand*\rtlola{\textsc{RTLola}\xspace}

\definecolor{CommentColor}{RGB}{42,0.0,255} 

\definecolor{bluekeywords}{rgb}{0.13, 0.13, 1}
\definecolor{greentypes}{rgb}{0, 0.5, 0}
\definecolor{redstrings}{RGB}{171, 114, 2}
\definecolor{graynumbers}{rgb}{0.5, 0.5, 0.5}
\definecolor{goldcomments}{rgb}{0.6, 0.4, 0.08}
\definecolor{monitorblue}{RGB}{18, 163, 38}

\lstdefinelanguage{Lola}{
  keywords=[0]{input, output, trigger, import, constant},
  keywordstyle=[0]\bfseries\color{bluekeywords},
  keywords=[1]{if, then, else, aggregate, defaults, offset, filter, hold},
  keywords=[2]{Int8, Int16, Int32, Int64, UInt8, UInt16, UInt32, UInt64, Bool, Float16, Float32, Float64, @1Hz, @2Hz, @5Hz, @10Hz, @100mHz, @1kHz},
  keywordstyle=[2]\color{greentypes},
  sensitive=false,
  comment=[l]{//},
  morecomment=[s]{/*}{*/},
  morestring=[b]',
  morestring=[b]"
}

\colorlet{eventcolor}{green!50!black}
\colorlet{periodiccolor}{blue!50!black}
\tikzstyle{event} = [draw=eventcolor, thin, fill opacity=.3, pattern=north west lines, pattern color=eventcolor]
\tikzstyle{periodic} = [draw=periodiccolor, thin, fill opacity=.3, pattern=north east lines, pattern color=periodiccolor]
\tikzstyle{signalname} = [text width=10em, minimum height=2em]
\tikzstyle{nameright} = [signalname, align=left]
\tikzstyle{namecenter} = [minimum height=2em, align=center, rotate=65]
\tikzstyle{nameleft} = [signalname, align=right]

\newcommand{\new}[1]{{#1}}

\begin{document}
\hyphenation{Time-stamp}
\hyphenation{Au-to-no-mous}

\title{Real-time Visualization of Stream-based Monitoring Data
  \thanks{
    This work was partially supported by the German Research Foundation (DFG) as part of the Collaborative Research Center Foundations of Perspicuous Software Systems (TRR 248, 389792660).
  }}

\author{
  Jan Baumeister\inst{1}\orcidID{0000-0002-8891-7483} \and
  Bernd Finkbeiner\inst{1}\orcidID{0000-0002-4280-8441} \and
  Stefan Gumhold\inst{2} \and
  Malte Schledjewski\inst{1}\orcidID{0000-0002-5221-9253}
}
\authorrunning{Baumeister et al.}
\titlerunning{Real-time Visualization of Stream-based Monitoring Data}
\institute{%
  CISPA Helmholtz Center for Information Security,
  66123 Saarbr\"ucken, Germany \\
  \email{\{jan.baumeister, finkbeiner, malte.schledjewski\}@cispa.de}
  \and
  Technische Universit\"at Dresden,
  01069 Dresden, Germany\\
  \email{stefan.gumhold@tu-dresden.de}
}
\maketitle

\begin{abstract}
  Stream-based runtime monitors are used in safety-critical
  applications such as Unmanned Aerial Systems (UAS) to compute
  comprehensive statistics and logical assessments of system health
  that provide the human operator with critical information in
  hand-over situations.  In such applications, a visual display of the monitoring data
  can be much more helpful than the
  textual alerts provided by a more traditional user interface.
  This visualization requires extensive real-time data processing, which includes
  the synchronization of data from different streams, filtering and
  aggregation, and priorization and management of user attention.  We
  present a visualization approach for the \rtlola monitoring
  framework. Our approach is based on the principle that the necessary data processing
  is the responsibility of the monitor itself, rather than the
  responsibility of some external visualization tool. We show how the
  various aspects of the data transformation can be described as \rtlola stream
  equations and linked to the visualization component through a bidirectional synchronous interface.
  In our experience, this approach leads to highly informative visualizations as well as to understandable and easily maintainable monitoring code.
  \donotshow{ \todo{adapt abstract} While monitoring is a
    great way to describe properties for CPS, the verdict of a runtime
    monitor is only as valuable as the reactions it elicits.
    Automated responses to the system are certainly desirable,
    however, in practice, a human supervises the system and initiates
    counter-measurements.  For this, the representation of the output
    of the monitor is as important as detecting incorrect behavior.

    Yet, this is particularly lacking for quantitative and stream-based monitoring techniques, where the output format rarely exceeds a textual representation.
    To this end, this tool paper presents a visualization component for the stream-based runtime monitoring framework \rtlola.
    The monitor evaluates input traces with respect to a specification.
    This specification includes two parts: one part describes the properties defining the correct behavior of the system and one part to prepare the data for the visualization tool, the focus of this paper.}
\end{abstract}

\section{Introduction}

Over the past decades, the scope of runtime verification has grown
from an essentially boolean check, indicating whether or not a program
execution satisfies a given formal specification, towards the
real-time computation of more and more expressive statistical data. A
typical example are Unmanned Aerial Systems (UAS), where the monitor
continuously collects and aggregates inputs from sensors and on-board
components to provide the human operator with critical information in
hand-over situations~\cite{uav2,rtutrv,rtlolacavindustrial}.

Traditionally, the interaction between runtime verification tools and
their users has largely been based on textual interfaces, such as
``alert'' messages that are issued in case of a violation of the
specification. In applications like UAS, however, such a simple user
interface is often no longer sufficient. In addition to understanding that
the monitor has detected a problem, the human operator must quickly
grasp the situation and decide on potentially time-critical corrective
action.  A well-designed \emph{visual} presentation of the available
data is therefore of critical importance for the safe operation of the
system.

\begin{figure}[t]
  \centering
  \includegraphics[width=0.95\textwidth]{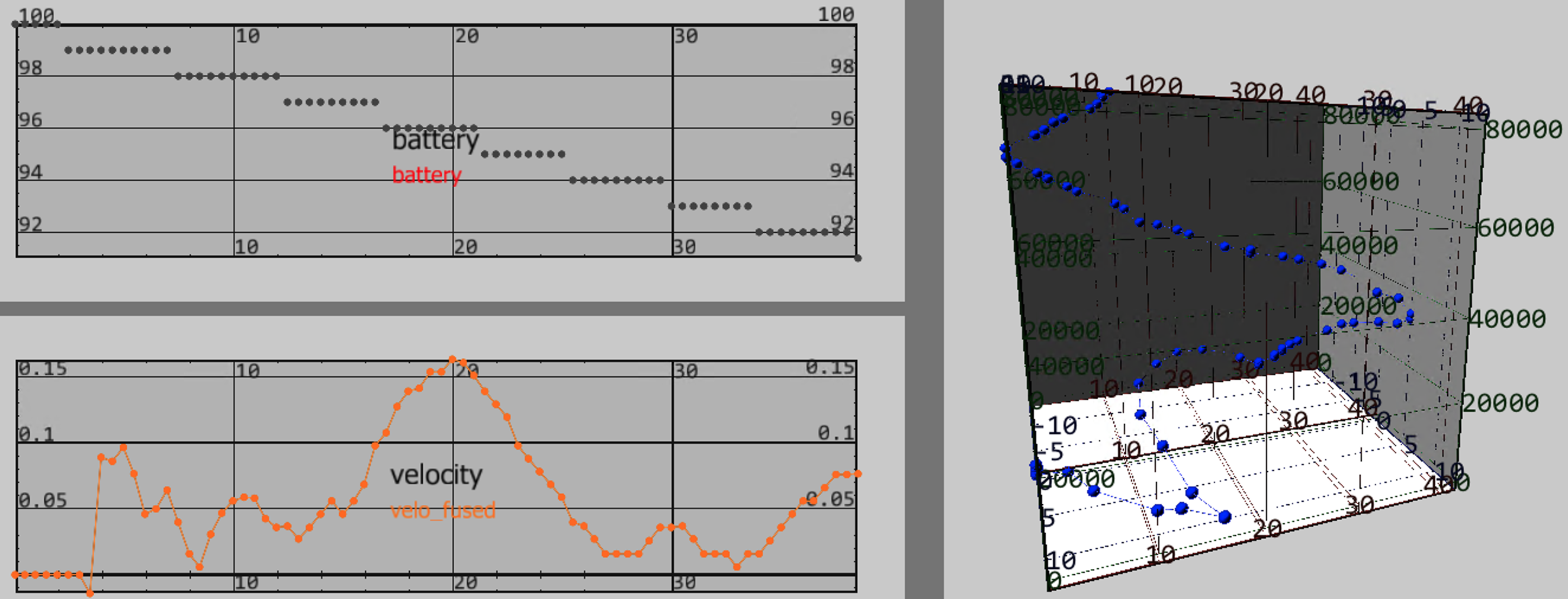}\smallskip
  \caption{\new{Screenshot of a visualization displaying the battery status, velocity, and GPS coordinates of a UAS.}}
  \label{fig:prototype}
\end{figure}

Generating useful visualizations is far from trivial.
First and foremost, the visualization must ensure that
important or dangerous information is clearly visible to the user;
because of the abundance of available data, data must
be prioritized, and less important data must be hidden in favor of
more important data. Similarly, the frequency of data points must be
adjusted to provide meaningful information avoiding overlaps and
adjusting for discrepancies in the availability of the raw data. All
these computations have to be adjusted in response to actions by the user,
who may look at multidimensional data from different angles or zoom
into data areas of particular interest.

In this paper, we report on our recent effort in extending the \rtlola
monitoring framework with real-time visualization
capabilities. \rtlola~\cite{rtlolacavtoolpaper,fpgalola} is a
stream-based monitoring framework for cyber-physical systems and
networks. \rtlola processes, evaluates, and aggregates streams of input
data, such as sensor readings, and provides a real-time analysis in
the form of comprehensive statistics and logical assessments of the
system's health. An \rtlola monitor is generated from a formal
description given in the \rtlola specification language. The
specifications consist of stream equations that translate input
streams into output streams. \rtlola specifications are statically
analyzed to determine the required memory and are then either directly
executed by the \rtlola interpreter, or compiled onto an FPGA.

The fundamental insight of our approach is that the data processing
needed to generate the visualization should
be the responsibility of the
monitor itself, rather than that of some external
visualization tool.  The monitor has access to all information and is
therefore in the best position to determine the relevancy of
individual data points. Because of the expressive power of the
monitoring language, the monitor also has the computational means to
interpolate and aggregate the raw data as required. Finally, keeping all data manipulations
in one place reduces redundancy and avoids errors and misinterpretations.

We organize the \rtlola specification of the data processing for the visualization into three functional areas:
{\bf 1. Data Synchronization:} This part of the specification guarantees \emph{synchronous} data updates for the visualization.
  This is important because the visualization combines data from different streams into a single entry in the plot: for example,
  a point indicating the position of a drone might be annotated with the speed of the drone.
  If different attributes have different timing, for example because of the variations in the frequency of the sensors, the monitor
  interpolates the missing data.
  {\bf 2. Filtering and aggregation:} This part of the specification avoids overlapping markers in the visualization, which are caused if readings arrive at a high rate.
  The monitor smoothes the input signal and adjusts the rate according to the current visualization.
  {\bf 3. Priorization and attention management:} This part of the specification determines the criticality of the available information and ensures that the human operator does not miss important information.

The monitor and the visualization component are connected via a synchronous interface.
The responsibility of the visualization component is to create the graphical display and to react to user requests.
Since this user interaction may affect the visibility of plots or change the scaling, a backchannel provides this information
to the monitor in the form of additional input streams.
\new{\cref{fig:prototype} shows a screenshot of our prototype implementation, which is based on the monitoring framework \rtlola and the visualization framework cgv~\cite{cgvRepo}.
  The monitor interacts with configurable 2D and 3D plots that support time-series plots, scatter plots, trajectory plots and multi-variate visualization through a flexible mapping of data attributes to the visual attributes color, opacity and size.}
We have applied our approach to the real-time visualization of UAS and other cyber-physical systems, based on existing \rtlola case studies; our experience suggests that adding the visualization specification inside the monitor leads to
highly informative visualizations as well as to understandable and easily maintainable monitoring code.

\subsection{Related Work}

This paper connects two traditionally separate areas of research: runtime monitoring and visualization. Somewhat surprisingly, visualization has not played a major role in monitoring research before. Despite a wide range of monitoring approaches, from formal logic~\cite{DBLP:conf/tacas/HavelundR02,DBLP:journals/fmsd/FinkbeinerS04,Donz2013,Raskin1997} to stream-based specification languages~\cite{lola,lola2, tessla, striver}, most tools have in common that they rely on textual, rather than visual, methods for data presentation.
This paper shows that stream-based monitoring languages like \rtlola are very well suited to carry out the needed data processing for useful visualizations.  Our focus on \rtlola is motivated by recent work on RTLola-based monitoring for UAS~\cite{rtlolacavindustrial, 10.1007/978-3-319-67531-2_3} and other cyber-physical systems~\cite{fpgalola, rtlolacavtoolpaper, DBLP:conf/tacas/BiewerFHKSS21}. However, the approach of the paper is clearly transferrable to other monitoring tools for CPS~\cite{copilot, DBLP:journals/fmsd/MoosbruggerRS17, LJBHCLR22}.

In the area of visualization, research on streaming visualization is also still in an early stage. A notable result are streaming processing models for data \cite{szewczyk_streaming_2011} and techniques for kernel density estimation in aggregated views over 2D maps \cite{lampe_interactive_2011}. There has been a systematic discussion of the suitability and problems of traditional visual analysis techniques~\cite{krstajic_visualization_2013,smestad2014interactive,dasgupta_human_2018}. Additionally, visualization frameworks~\cite{fischer_real-time_2012} and visualization techniques that
allow the user to better cope with changes over time have been developed, such as zoomable navigation~\cite{7043852}, paged views~\cite{fischer_nstreamaware_2014}, and transformation-based smooth transitions~\cite{li_streammap_2018}.
These approaches differ substantially from the approach taken in this paper, in that these are independent visualization tools, which prepare the data for the visualization independently of the monitor. By contrast, our setup
tightly integrates the monitor with the visualization, keeping all data manipulations in one place.

\section{RTLola}
\label{sec:rtlola}

\rtlola is a
stream-based monitoring framework for cyber-physical systems and
networks. An \rtlola monitor is generated from a formal
specification description given in the \rtlola specification language.
The specification consists of stream equations that describe the transformation of incoming data streams into output streams, and a set of trigger conditions that result in notifications to the user. 
The \rtlola framework includes automatic static analysis methods that ensure the predictability of the monitor with respect to memory consumption and other relevant properties. 
We illustrate the \rtlola specification language with a small example; for more details, we refer the reader to \cite{rtlolacavtoolpaper,fpgalola}.
\begin{lstlisting}
input gps: (Float64, Float64), charge: Float64, time: Float64
output charge_time @charge := time.hold(or: 0.0)
output filtered_gps filter gps != (0.0,0.0) := gps
trigger $\delta(\texttt{charge})$ / $\delta(\texttt{charge_time})$ > 2.0
trigger filtered_gps.0 > 6.0 $\land$ filtered_gps.1 > 6.0
\end{lstlisting}
The specification declares three input streams:
The first stream \lstinline|gps| represents readings received by the GNSS (global navigation satellite system) module, the second stream \lstinline|charge| shows the battery charge status, and the third one \lstinline|time| contains the current time.
Next, the specification declares the \lstinline|charge_time| output stream, which filters the \lstinline|time| stream to timestamps of newly received battery readings.
For this, it binds the timing of the \lstinline|charge_time| stream to the timing of \lstinline|charge|, indicated by the \lstinline|@charge| annotation.
In \rtlola, such a filter is called a static filter.
As \lstinline|time| might have a different timing, the value is accessed via a 0-order hold interpolation.
The next stream \lstinline|filtered_gps| uses a dynamic filtering to exclude noisy sensor readings.
In our example, the GNSS sends $(0.0,0.0)$ coordinates during initialization that the specification can discard.
The last two lines contain triggers checking if there is an unusual drop in the battery status and if the coordinates do not exceed some thresholds.

In \rtlola, the static and dynamic filters are combined into a \emph{pacing type}, which defines the timing of each stream.
This type is either inferred or explicitly annotated.
\rtlola's type checker verifies the timing of the streams and \rtlola provides different operators to interpolate data if the timing constraints cannot be guaranteed in the stream expression.

\donotshow{
	\begin{itemize}
		\item stream-based specification language
		\item successfully used to describe correct behavior of a system [CAV case study]
		\item RtLola by example:
		      \begin{itemize}
			      \item different types of streams
			      \item pacing: asynchronity (static, example interpolation) and dynamic filtering
			      \item real-time properties + aggregation
		      \end{itemize}
	\end{itemize}
}

\section{Generating Visualization Data}\label{sec:contribution}
\donotshow{
	\begin{itemize}
		\item Motivation: trigger bad need some kind of visualization with human in the loop
		\item Concept with \cref{fig:overview}: Combination between monitor and visualization, therefore specification gets data from visualization and produces data for visualization
		      \subitem first which data STREAMS need visualization and in which plot combination
		      \subitem Concrete which new output streams are now added to the spec
		      \subitem Concrete which input streams with motivation of backchannel to describe current context
		\item 3 steps to define the new streams:
		      \begin{itemize}
			      \item Data synchronization: plots need data synchronously, not always the case with sensors, but we not a new problem in monitoring
			      \item filtering and aggregation: dependent on the context different rate to plot data (overlapping)
			      \item priority: some information more important, monitor has this context of priority and this can be applied to visualization
		      \end{itemize}
	\end{itemize}
}
%
%

We now describe in more detail the generation of visualization data with \rtlola stream equations. 
For the communication from the monitor to the visualization component, the specification contains output streams that are mapped to plot coordinates and visual attributes such as size and color.
It also contains an output stream per axis, setting its displayed range.
For the backchannel, 
the specification has input streams that receive the data from the visualization reflecting the interaction of the visualization component with the user. For each 2D-plot, we include one input stream to transfer the current scale factors; for each 3D-plot we include two input streams, representing the projection matrix and window size.
Additionally, the specification has an input stream for each plot indicating which plot is visible. 

We structure the generation of the visualization data into three areas:
\emph{Data Synchronization and Interpolation}, \emph{Filtering and Aggregation}, and \emph{Prioritization}.
%
\begin{figure}[t]
	\input{specification}
	\caption{\rtlola specification demonstrating the interplay.}
	\label{fig:specification}
\end{figure}
\new{For each area, we shortly describe the problem, then describe the mechanism of how \rtlola solves the task and explain the solution in more detail with our running example shown in \cref{fig:specification}.}
We display the coordinates of a GNSS in a 2-dimensional plot, and the remaining battery charge is mapped onto the color of the marker.
The input streams \lstinline!charge! and \lstinline!gps! represent the sensor readings followed by the streams \lstinline!pixel_scale! and \lstinline!visible! implementing the backchannel.
The output streams \lstinline!xLim! and \lstinline!yLim! compute the upper and lower bound per axis which is in our case the global minimum and maximum.
Alternatively, our plot could represent with the following stream expression the data over a time-period $\sigma$:
\begin{lstlisting}
output x_limits: (Float64, Float64) @1Hz
  := (gps.0.aggregate(over: $\sigma$, using: min), gps.0.aggregate(over: $\sigma$, using: max))
\end{lstlisting}
The next streams are helper streams to filter the data and the last output stream \lstinline!marker! contains all information needed for a new marker in the plot.

\subsection{Data Synchronization and Interpolation}
\label{sec:contribution:dataSyn}
\donotshow{
	\begin{itemize}
		\item Data in one plot need to be synchronized \checkmark
		\item In practice not always the case, e.g. different sensors have different frequencies \checkmark
		\item Example malte \checkmark
		\item Same problem in monitoring when arguing about different sensor values \checkmark
		\item in rtlola we have the concept of pacing types \checkmark
		\item what does this mean for our output streams: \checkmark
		      \subitem give a plot or subplot a pacing and annotate the corresponding outputstreams with this pacing\checkmark
		      \subitem type checker guarantees that every value is accessable\checkmark
		      \subitem reminder: iterpolation, 0-order hold to get data from different timing\checkmark
		      \subitem back to our example: explain on example \checkmark
	\end{itemize}
}
For drawing a marker, the visualization needs to know all its visual attributes which might be based on sensors with different frequencies. 
We use \rtlola's type system to guarantee that the monitor sends synchronized updates per plot.
%
%
%
%
For this task, we use the concept of pacing types as introduced in \cref{sec:rtlola}.
We define a pacing for each plot and annotate the streams for this plot with the desired pacing.
Then, we use \rtlola's type checker to verify that the data is available.
In our example, we want to create a marker whenever at least one sensor sends an update and therefore use the disjunction of the two input streams as the pacing type.
This pacing type \lstinline|@gps$\lor$charge| is annotated to the stream \lstinline!marker!.
Similarly, we want to update the axis limits, represented by the streams \lstinline!xLim! and \lstinline!yLim!, whenever we get a new GPS sensor reading.

We cannot directly access the current value of each stream because they may have different timings. Instead, we specify how missing data is interpolated.
\rtlola offers different approaches for this task, e.g., by using data aggregations, zero-order hold operations, or even different forms of data interpolations.
In our example, we use a zero-order hold on the missing data.

\subsection{Data Filtering and Aggregation}
\label{sec:contribution:filter}

\begin{figure}[t]
	\begin{minipage}{\linewidth}
		\begin{minipage}{0.47\linewidth}
			\centering
			\includegraphics[width=0.65\textwidth]{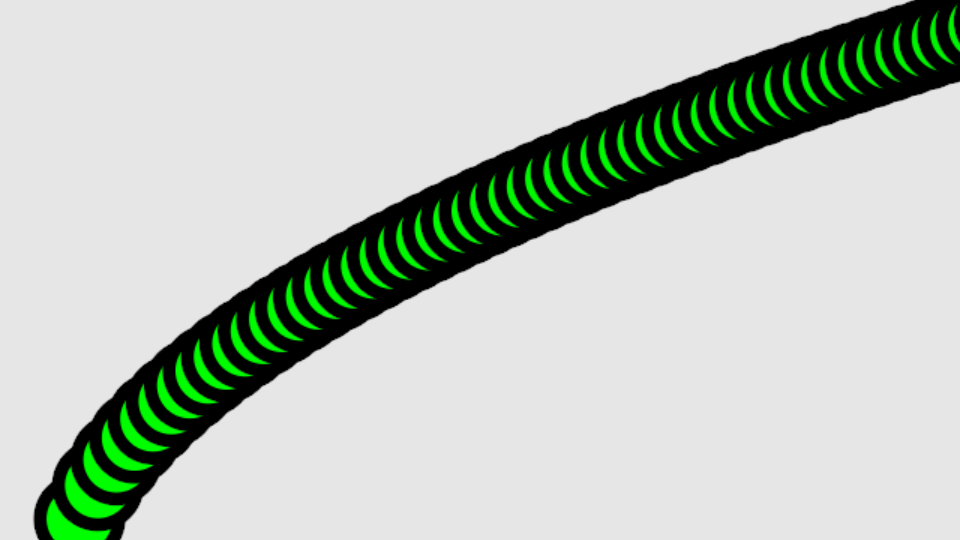}
		\end{minipage}
		\hfill
		\begin{minipage}{0.47\linewidth}
			\centering
			\includegraphics[width=0.65\textwidth]{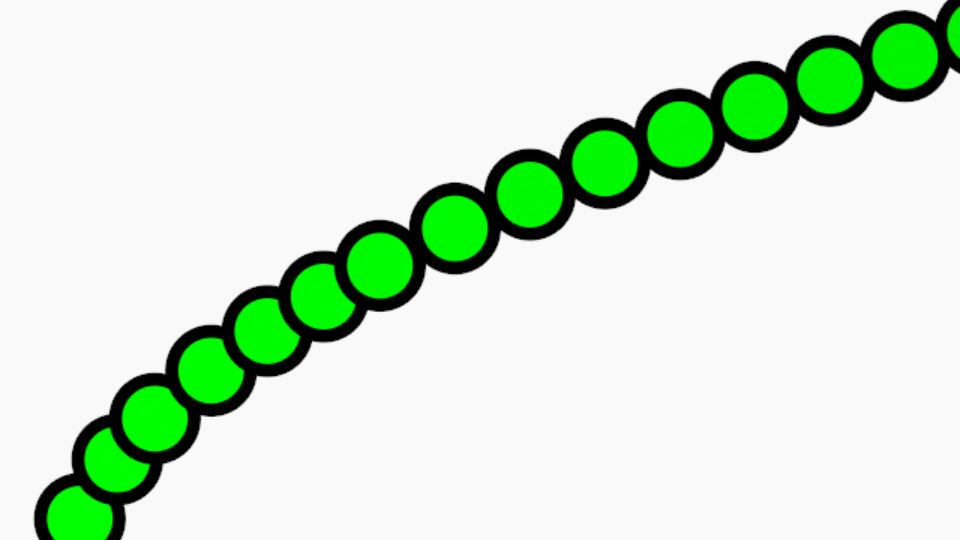}
		\end{minipage}
	\end{minipage}
	\caption{Screenshot of the prototype with different monitors. On the left side, we use a monitor forwarding all data, whereas the right monitor filters the data using the specification in \cref{fig:specification}.
	}
	\label{fig:overlap}
\end{figure}
This section shows how a monitor prepares data to provide more understandable updates to the user.
\cref{fig:overlap} shows two plots from the same execution.
On the left side, the monitor forwards all data to the visualization resulting in overlapping markers.
These overlapping markers overload the plot with unrequired information and even overlap some information, as in our example, the color illustrating the battery status.
On the right side, the monitor tailors the data for visualization and transfers the prepared data, so we do not have this problem.

\new{The monitor on the right uses \rtlola's dynamic filtering mechanism to prepare the data.
With this filtering approach, the monitor can dynamically adapt the throughput to the visualization.
Dependent on the scenario, different filters are helpful:
}
For example, a filter forwards the GPS data dependent on the current velocity or increasing the sample rate if the system violates a property could be easily expressed in \rtlola.
In our running example, we forward the markers only if the difference between the new marker and the previous marker exceeds a threshold and if the plot is visible.
For this, the output stream \lstinline!marker! has two such filters connected by a disjunction.
Both filters use the information provided by the visualization to describe the current state of the plot -- visibility and scaling.
With the second disjunct, we ensure that a marker is transmitted only if the plot is visible.
The other filter is encoded by the stream \lstinline!send!.
In this stream, we decide whether the difference to the previously transmitted marker is sufficient to warrant a new marker.
For this, we first compute the distance between the pixel coordinates of the candidate marker and the last marker based on the current bounds and scaling, and compare it with a defined threshold that prevents overlapping.
We also check whether the difference in the charging level warrants a new marker.
\new{A similar approach also applies to 3D plots. Instead of the \lstinline|pixel_scale|, we use the projecting matrix that encodes besides the scaling, the viewing rotation, and perspective.
}

Depending on the scenario and the size of the plot, it can be useful to aggregate values (such as by computing the average, minimum, and maximum of the data since the last marker) instead of dropping values.
In \rtlola, this can easily be done using the corresponding aggregation functions.

\subsection{Attention Management}
\label{sec:contribution:attention}
\begin{figure}[t]
	\begin{minipage}{\linewidth}
		\begin{minipage}{0.47\linewidth}
			\centering
			\includegraphics[width=0.65\textwidth]{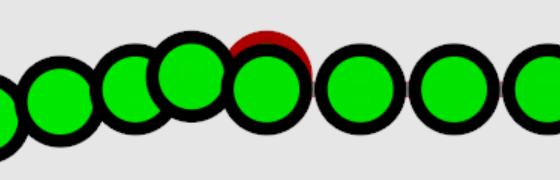}
		\end{minipage}
		\hfill
		\begin{minipage}{0.47\linewidth}
			\centering
			\includegraphics[width=0.65\textwidth]{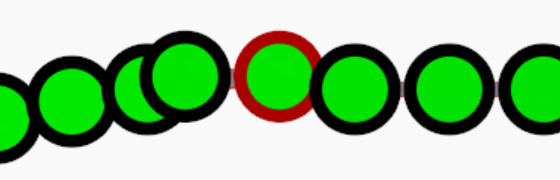}
		\end{minipage}
	\end{minipage}
	\caption{On the left side, one marker with irrelevant information covers a critical marker, whereas the monitor on the right encodes a form of priority.}
	\label{fig:attentionManagement}
\end{figure}

In \cref{sec:contribution:filter}, we have already discussed how the monitor can filter data points if they do not contain relevant information.
Often, however, this is not sufficient, and we need to \emph{prioritize} information:
\cref{fig:attentionManagement} shows two plots containing a critical state that the operator should recognize, illustrated by the red marker.
This marker is partially covered on the left by plotting a new marker that does not contain this information anymore.
The operator can easily miss this information, so 
the monitor on the right prioritizes them and thus does not send the candidate marker to the visualization tool.

\new{Again, we use \rtlola's dynamic filtering mechanism to prevent the coverage of higher prioritized information.
We also introduce new streams encoding the priority of information and checking the coverage of markers. 
}
In our running example, we need to change the stream expression of \lstinline|send| 
and add the following streams to the specification: 
\begin{lstlisting}
	output critical: Bool @gps$\lor$charge := $\dots$
	output marker_lc @gps$\lor$charge
	:= if send $\land$ critical then marker else marker_lc.offset(by:-1, or:$marker_s$)
	output $\delta$xc @gps$\lor$charge
	:= (gps.hold(or:$gps_{s}$).0 - marker_lc.offset(by:-1, or:$marker_{s}$).0) / (xLim.hold(or:1.0).1-xLim.hold(or:1.0).0) * pixel_scale.hold(or:$pixel\_scale_{s}$)
	output $\delta$yc @gps$\lor$charge := $\dots$
	output send @gps $\lor$ charge
	:= (sqrt($\delta$x**2.0 + $\delta$y**2.0) > $\tau_{gps}$ $\lor$ $\delta$c > $\tau_{charge}$) $\land$ ((sqrt($\delta$xc**2.0 + $\delta$yc**2.0) > $\tau_{c}$ $\lor$ critical)
\end{lstlisting}
\new{
The stream \lstinline|citical| encodes the priority of a marker
and the next stream \lstinline|marker_lc| stores the values of the last critical marker.
The change stream \lstinline|send| now also determines if a potential new marker would overlap the last critical marker by computing the distance between the markers with the help of \lstinline|$\delta$xc| and \lstinline|$\delta$yc| and then checking if this distance is sufficient.

}

%
Preventing covering markers with less relevant information is only one example of how we can encode the priority of information.
Another example occurs when the specification aggregates data points to make the plots more readable:
With aggregations, we lose information about the system.
In general, this behavior is intended because the human supervisor cannot process all information from every sensor.
In critical situations, however, the operator usually is focused on the part that misbehaves.
In these situations, the monitor can switch to transferring each data point instead of aggregating them, or it might reduce the required difference for new markers, so the supervisor gets all the required information.
Such a property can be expressed in \rtlola by adapting the timing of a stream or by using different aggregation functions.

\section{Conclusions}
\label{sec:conclusion}
\new{In this paper, we have introduced a principled approach to the real-time visualization of stream-based monitoring data. The key contributions are the novel design principle, which shifts the responsibility for the data preparation from the visualization component to the monitor; the organization of the approach into 
into three major functional areas; and the solution of the visualization challenges with the mechanisms of a stream-based monitoring language. 
}
%
%

Our practical experience with the approach of the paper has been very
positive.  We have used the approach to visualize stream-based
monitoring data from recent case studies that use \rtlola for
UAS~\cite{rtlolacavindustrial, 10.1007/978-3-319-67531-2_3} and other
cyber-physical systems~\cite{fpgalola, rtlolacavtoolpaper,
  DBLP:conf/tacas/BiewerFHKSS21}. The visual tools provided by cgv
have proven very useful for the type of data produced by our
monitors. For example, we have visualized the failure of the GPS
module in a drone, which was recognized by the system health check in
the existing monitor specification, by adding a halo to the markers of
the estimated position and by increasing the marker frequency.  While
clearly more research is needed in order to determine the best
visualizations, our experience already indicates that this type of visualization is very helpful in quickly
understanding complicated situations.

We hope that this paper will inspire other developers of runtime verification tools to invest in real-time visualization as well. We believe that our ``monitoring-oriented'' visualization approach provides a significant step towards meaningful visualizations that exploit the wealth of information available within the monitor. In future work, it might even be possible to integrate explicit visualization operators into monitoring languages like \rtlola, and thus largely automate the visualization process presented in this paper.

\bibliographystyle{splncs04}
\bibliography{bibliography}


\end{document}